# Porous Semiconductors: Growth and Applications

Colm O'Dwyer[1,2,3]

[1] School of Chemistry, University College Cork, Cork, T12 YN60, Ireland
[2] Environmental Research Institute, University College Cork, Cork T23 XE10, Ireland
[3] AMBER Research Center, Trinity College Dublin, Dublin 2, Ireland

Big pores, small pores, ordered pores, random pores – they all have a function and as is often found, show behaviour in new materials that is not always predicted or obvious at the outset. I started my research journey trying to put extremely thin films onto near-perfect III-V crystals to control (opto)electronic properties and when the first TEM on our campus showed remarkable pore growth and structure in InP almost 21 years ago, the electrochemical modification of the InP made more sense. In this paper, I will summarise a few aspects of research into porous materials and semiconductors, from porous InP that led to studies of other porous semiconductors such as silicon, GaN, ZnO and Indium Tin oxide (ITO), to periodically ordered photonic crystal porous structures and some optical, thermal and electrochemical properties, photocatalysis, studies in batteries and related that were enabled or modified by the porous structure.

## Introduction

Porous semiconductors continue to receive considerable attention because of functionality afforded by random or ordered structuring on the nano or mesoscale, and the litany of discoveries and applications have shown that control of porosity continuously gives useful and sometimes surprising practical benefits and applications. For instance, a wide range of mono- and compound semiconductors can be rendered porous through electroless or electrochemical etching[1-26], and such top down approaches allow a high degree of control over porosity formation. Additionally, the resulting skeleton formed through these means allows the possibility for arrays or arrangements of nanostructured materials such as nanowires (NWs), mesoporous materials, nanorods, photonic crystals and many other structural forms. When the sizes of the remaining semiconductors is reduced, photonic and phononic confinement effects can alter (electro)chemical, mechanical, and other physical characteristics.

Rechargeable batteries have been critical for enabling portable consumer electronics and are beginning to be used in electric vehicles. Current lithium-ion battery anode research involves significant investigations of semiconducting and metalloid materials, particularly Si as its theoretical specific capacity is >4000 mAh/g under deep cycling conditions. Previous theoretical studies showed that porous Si with a large pore size and high porosity can maintain its structure after Li ion induced alloying and swelling. Silicon has maintained a strong fundamental and applied research value, and recently has become one of the most significant materials, when structured on the nanoscale, for Li-ion battery anodes[27-29] and in thermoelectric materials evaluation[30-32].

Transparent conducting oxides are also important for electronic and photonic devices. The trade-off between their conductivity and transparency[33] is a fundamental limiting factor, which can be overcome to some extent through controlled top-down or bottom up porosity formation. Researchers have sought to control shape, aspect ratio and crystalline arrangement[34,35], and recent improvements in synthetic methods[36-38] have led to the direct integration of functional nanostructures into nanoscale devices. Indium tin oxide (ITO)[39,40] is the most important transparent conducting oxide and is therefore used in a wide range of applications, and many advances have been made in complimentary applications of another compounds semiconductor, ZnO, as a wide bandgap oxide for oxide-based electronics.

Porosity has been the key enabler for many of the discoveries and applications for several semiconductor materials from electronics and photonics to energy storage, catalysis and batteries, and many more besides. This paper will focus on semiconducting materials in porous form and the investigations we have been doing across many areas of technical interest to ECS and beyond. These include Si, GaN, InP, ITO, ZnO, many metal oxides and also some 3D printed structures.

**Results and Discussion**

*Porous InP*

Much of the early research on porosity in compound semiconductors has been on gallium arsenide and indium phosphide, inspired by the advances in controlled etching of compound semiconductors but also from progress in the formation of porous silicon[41-48] and the possibility of quantum confinement-induced light emission from silicon. It has been shown that pore growth is affected by electrolyte concentration[18-25] as well as substrate type, orientation and doping density. Anodic etching of n-type semiconductors does not readily occur in the dark since it requires holes in the valence band. At highly positive potentials, however, tunneling of carriers across the depletion layer is sufficiently highly doped wafers can occur and the resulting holes in the valence band enable etching to occur and this effect allows a wide range of porous structures to be formed and controlled electrochemically.

Anodic etching protocols are now more common and our work complimented observations on other III-V porous semiconductors by examining the mechanism of pore growth, propagation and branching in (100)-oriented n-type InP wafers in strongly basic electrolyte such as KOH[49]. The more common practice in etching up to then used acidic electrolytes.

Anodization of highly doped n-InP in aqueous KOH results in the formation of a nanoporous sub-surface region beneath a thin (typically ~40 nm) dense near-surface layer. Linear sweep voltammetry showed a pronounced anodic peak corresponding to the formation of the porous region. Porosity in InP begins by the formation of surface pits the penetrate the surface. These were confirmed by AFM and by cross-sectional TEM as shown in the overview in Fig. 1. The surprising aspect that allowed a good mechanistic insight into the details of pore growth and formation was the observation of very well-defined domains of pores beneath surface. They are separated from the surface by a thin ~40 nm layer, which is similar to the depletion region width for this semiconductor carrier concentration and potential. One of these domains forms under every pit and their merger is what creates the full porous layer. While our observations over the years followed the random pit formation process, controlling site specific domain formation and merger can in principle be controlled by creating surface-bound localised regions where surface pits occur. Figure 1 shows the overall process, tracked from a surface pit right through to the final porous layer.

The investigations over the years for porous InP focused on the mechanism, as there were several open questions for anodic etching and pore growth in III-Vs in general, and the uniqueness of pore growth on n-InP in KOH. We developed a model of porous structure growth that is based on pore propagation along <111>A directions.[50-52] In a series of detailed analyses, this model accurately predicted a porous domain with a specific truncated tetrahedral shape. This is characterized by a structure composed of trapezium-, triangle- and square-shaped cross-sections when viewed in the (011), (011) and (100) planes, respectively. Examples of the overall structural model and TEM evidence in also shown in Fig. 1, and more details in associated references. Pores are cylindrical and have well-developed facets only near their tips. Each pore tip is the apex of a pyramid formed by three <111>A facets. Pore growth occurs only near the tips, where the electric field is higher.

These porous InP structures, and related structures in GaAs and other compound semiconductors are complex but offer potential opportunities in photonics if the nucleation, growth and pore growth, merging and domain structure can be controlled. Possibilities include modified scattering, graded refractive index surface layers, photonic band gap tuning and other optical effects caused by either random or ordered arrays or pores in a single crystal semiconductor.

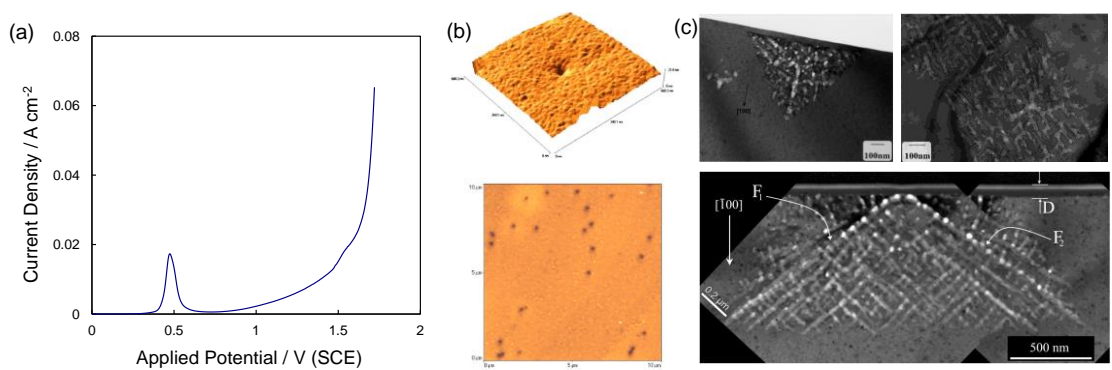

**Figure 1.** (a) Linear sweep voltammogram of an n-InP(100) electrode in 5 mol dm$^{-3}$ KOH. The current peak at ~0.5 V is where pitting, pore growth and porous layer formation occurs. (b) AFM images of the onset of pore growth, where pits are formed on the surface of the InP. (c) Cross-sectional and plan view images of a porous domain, which forms under a near surface layer ~40 nm thick and etches into the crystal from the surface pit. Each porous domain is a truncated tetrahedral geometry and a full layer is formed by merging or these domains.

*Porous Silicon*

Porous silicon was one of the first porous semiconductors that showed promise in optoelectronics and related fields. The discoveries of Canham, Lehman and Föll showed how quantum wires of silicon could allow quantum confinement and thus direct red light emission from what is normally an indirect bandgap semiconductor.[41-43] While many electrochemical methods followed, a simpler approach based on electroless etching, more commonly referred to as metal assisted chemical etching (MACE) was reported. This approach is useful, does not require large current (for large area wafers) potentio/galvanostats and allows for large-scale, and rapid fabrication of high-quality[53], well-aligned vertically oriented Si NWs. Like InP, semiconducting materials where pores form can be controlled the etching bath, additives, temperature but also substrate doping type and doping concentration. Once pores could be formed in silicon in many forms and from many ways, their applications were numerous, spanning Li-ion batteries[53-62], thermoelectrics[63,64], solar cells[65], water oxidation[5] and electroluminescent devices.

The etching mechanism for MACE is mostly commonly accomplished using Ag deposition or Ag nanoparticles on the surface of the silicon. Ag is oxidized into $Ag^+$ ions by $H_2O_2$ in the etching bath[66,67] and the $Ag^+$ ions can be reduced during further Si oxidation and etching through nucleophilic attack. Electroless hole injection drives the process. Usefully for many applications, the pores in silicon can be made at multiple length scales. A popular result is the formation of nanowire (NW) mats or layers and offers a nice top down approach compared to bottom-up VLS-type growth. MACE can result in high sidewall roughness, and the control of this surface roughness and the development of internal mesoporosity[68,69] within each NW are key challenges for reproducible large scale formation of functional nanoscale Si. On the other hand, this multiscale porosity that forms NWs from a single crystal, and internal mesoporous within each NW offers a very high semiconducting surface area structure that has been exploited for thermoelectrics and optical emission. The final Si nanostructure generated can be controlled by the substrate doping type level.[70-75]

We studied MACE of Si NWs and found that internal mesoporosity can be controlled by doping density, which in a certain electrolyte defines the depletion layer width and the dimension of the internal walls of the mesoporous structure.[76-78] Control of doping density, conductivity type and resistivity can also allow roughened NWs with solid cores and more complex porous layers of NWs that grow at specific angles to the substrate from (111)-oriented Si wafers (see Figure 2). Recently, it was found that (100) wafers with resistivity of 6-8 Ω·cm relieve vertical <100> NWs at a relatively low volumetric ratio of $HF/H_2O_2$ of 3:1 while <111> NWs are generated when the HF concentration is increased.

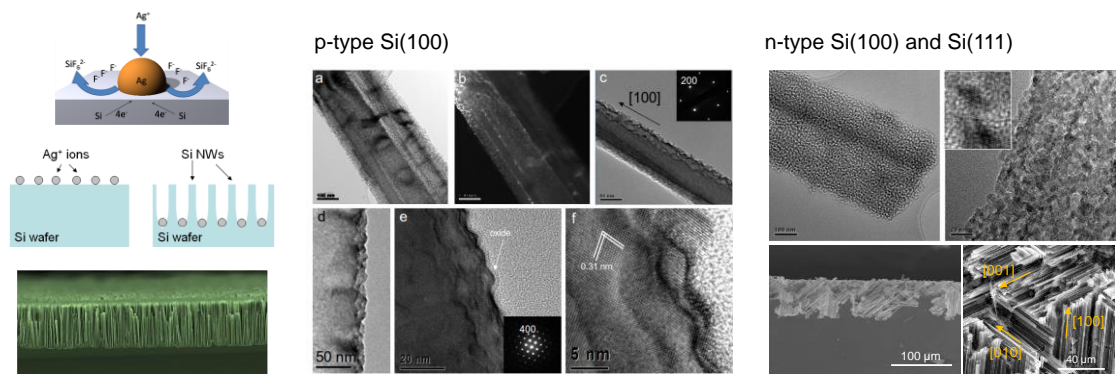

**Figure 2.** Representation of metal-assisted chemical etching of silicon using $Ag^+$ ions in a $H_2O_2$/HF solution to form a Si NW porous layer. For p-type Si(100), the NWs are solid core with roughened outer walls. n-type Si(100) also forms a NW layer, but the internal structure of the NW is mesoporous. For Si(111) substrates, etching proceeds also along <100> directions, creating a more complex NW layer where domains of NWs are aligned at ~57° to other domains of NWs.

These mesoporous Si NW layers are also intense red light emitters. We reported an investigation of photoluminescence from these highly porous structures. Mesoporous $n^+$-Si(100) nanowires and nanocrystal-decorated rough surface p-Si NWs formed by MACE showed time-resolved photoluminescence (PL) measurements indicated long (tens of μs) radiative recombination lifetimes. The red luminescence at ~1.7 eV is visible with the naked eye, as shown in Fig. 3. $n^+$-Si NWs internal mesoporosity is similar to a labyrinthine network of silicon nanocrystals in a random mesoporous structure, very different to the p-type Si(100) NWs which have a rough surface and a solid NW core. The rough surface comprises various orientations of Si nanocrystals that also show red luminescence ~1.8 eV, respectively.[79]

Finally, the roughness and degree of porosity affect temperature-dependent propagation of phonons. In order to realize increasingly efficient thermoelectric materials and devices, various methods have been used to suppress with phonon propagation and heat conduction.[80-83] Several approaches exists such as the addition of impurities, altering of crystal structure[84], or by modification to an alloy compounds with grian boundaries to scatter phonons and lower thermal conductivity and diffusivity.[85] Nanowire (NW) structures are also promising for thermoelectrics.[86,87] Roughened Si NWs produced by electroless etching of Si wafers exhibited severely suppressed thermal conductivity as low as 1.6 W m$^{-1}$ K$^{-1}$ for a 56 nm diameter NW at 300 K.[64] We recently showed that MAC-etched Si NWs directly from p-type Si forms layers of NWs with polycrystalline roughness on the outer walls of the NW made up of grain boundaries that are ~ 2-5 nm apart, providing phonon scattering capability.[76] Figure 3 shows the NW surface roughness and the nanocrystalline features. By using temperature-dependent Raman scattering, we showed that these MACE NWs require 4-phonon processes to explain the Raman scattering shifts.[88,89] The thermal conductivity of the Si NWs was calculated using the same method, with the local temperature calculated using the Raman shift of the zone-center LO phonon. At a laser power of 25 mW under 532 nm excitation, the LO mode shifted to a value of 513.8 cm$^{-1}$, corresponding to a local temperature of 600 K. This gives an estimate of the thermal to be in the range 12-18 W mK$^{-1}$.

The nanostructure of the roughness features and evidence of multi-phonon processes required to explain optical phonon mode shifts and asymmetric broadening with increasing temperature, may prove to be useful in explaining the uniquely suppressed thermal conductivities observed in Si NWs that are not coherent nor smooth. Surface roughness alone and the spectrum of roughness wavelengths may not be the sole basis for the several orders of magnitude reduction in thermal conductivity.

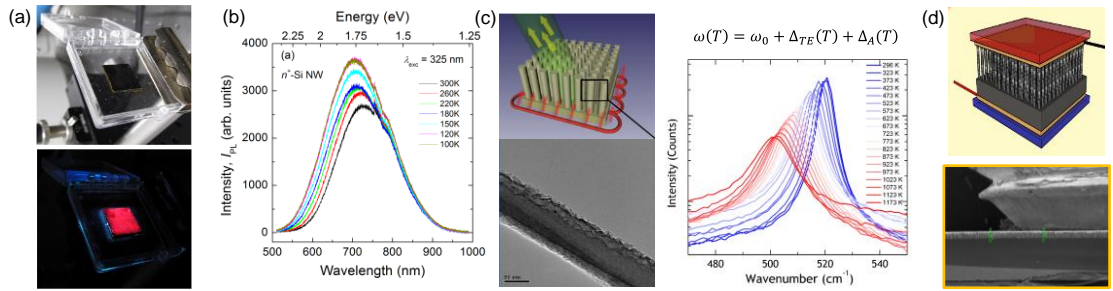

**Figure 3.** Pictures of a 1 cm$^2$ coupon of silicon with MACE Si NWs. The picture underneath shows intense red photoluminescence from the NW layer under optical excitation at 385 nm, with corresponding PL spectra form 100 – 300 K shown in (b). (c) A single p-Si rough wall NW and the Raman scattering spectra form the p-Si NW layer from 296 – 1173 K showing LO phonon shift. (d) Schematic of the thermal interface material made from a Si NW layer.

*Porous GaN*

Creating pores in III-N materials is conventionally more difficult than in many other semiconductors. Nitride compound semiconductors, particular those of wide bandgap, require electrochemical conditions that allow favourable charge transfer to drive an etching process, but also an electrolyte that facilities the decomposition of robust compounds that are relatively inert in acidic and basic solutions. Typically, generation of valence band holes that aid the dissolution of the III-N into a porous structure is more easily achieved under photoelectrochemical conditions using high energy photons near-resonant with the bandgap energy of the nitride semiconductor.[90-93] Many years of research were devoted to etching of GaN and related compounds during the development of devices beyond the remarkably successful white light and blue light LED breakthroughs made using these wide bandgap materials.[94-100]

Etching porosity into GaN for example, is a top-down process and starts with the benefit of having a high quality epitaxial GaN layer. This quality of GaN is also a prime reason etching this stable compound can be more difficult that other semiconductors, so alternative ways of producing porous GaN have been explored. Notable among them is chemical vapor deposition (CVD), which can create complex porous GaN without post growth etching procedures. Some years ago, with colleagues at URV in Tarragona, Spain, we investigated CVD of porous structures of GaN using this bottom-up growth method. In this approach, CVD growth initiates GaN seed growth that form distorted hour-glass shaped structures. Porosity is initiated and propagates on the (0001) face of the material in each structure. The entire porous layer is then formed by many individual crystalline structures randomly oriented in the layer to give a highly porous structure, as shown in Fig. 4. Using this technique, we have demonstrated low resistivity Pt and Au ohmic contacts to porous *n*-type GaN. This type of contact of such work function metals to wide bandgap GaN was made possible by the formation of intermetallic seed layer from a vapor-solid-solid (VSS) mechanism.[101]

One important aspect of bottom-up growth of GaN is to ensure high purity and also control of carrier concentration and conductivity type. GaN by CVD is unintentionally n-type, but we also demonstrated *p*-type porous GaN by doping it with Mg from a $Mg_2N_3$ precursor during CVD to give a carrier concentration approaching, ~$10^{18}$ cm$^{-3}$.[102,103] By changing the concentration of $Mg_2N_3$ to excess, CVD can form a polycrystalline high-κ oxide on the GaN surface under the ohmic metallic Mg-Ga alloy interlayer contact. A clean interface is kept between each layer and this one-step method allowed the fabrication of porous GaN-based MOS-type diodes on silicon substrates.[104]

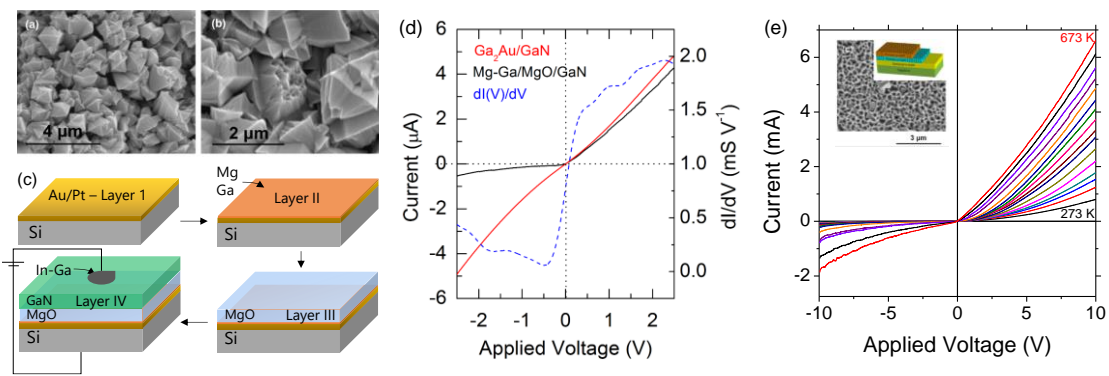

**Figure 4.** SEM image of porous GaN particles grown on Si(100) substrates coated with (a) a 20 nm film of Au and (b) a 20 nm film of Pt. (c) Mechanism of the creation of MgO/GaN layers of the MOS diode by subsequent catalytic VSS growth during CVD. (d) I-V and dI/dV curves for porous n-GaN with a MgO dielectric layer, and from a Au-seeded porous GaN layer on silicon. (e) I-V curves of the porous n-p GaN diode recorded at T = 298-673 K.

Porous materials, particularly optical and electronic materials such as semiconductors usually require very high quality and controlled growth of well-defined thin films or structured surface features.[105] Random porosity can tend to be difficult in these applications as the energies, smoothness, doping, interfacial species and more can affect optical and electrical characteristics profoundly. By changing the substrate, it was possible to grow porous GaN[106,107] layers by CVD in two steps to form porous GaN rectifying p-n junctions that exhibit stable rectification in the ±1-5 V range.[108] LEDs fabricated using porous GaN are possible using p-n junction diodes formed by an *n*-type porous GaN layer deposited on a *p*-type non-porous GaN substrate, as shown in Fig. 4.[107,109] While further development is required for porous semiconductor optoelectronic devices, this proof of concept show the potential for porous semiconductors to be a platform for sensing, electrochemical water splitting and other technologies where high surface areas are required together with good control over electrical and optical characteristics. Additionally, extension to InN and AlN and related alloys may prove useful for graded index or porous white light LEDs with reduced output reflection losses from reduced impedance matching losses at the interface with highest index contrast.

*Porous ZnO*

Zinc oxide (ZnO) has been one of the most investigated wide bandgap semiconductors in recent decades.[110-121] A lot of this popularity is due to the relatively simple synthesis conditions needed to make a very wide range of structures and films of this material from solution.[122-126] While such a amenable protocol has allowed access to this materials to many researchers around the globe, the inherent electronic structure of ZnO and the influence of defects on many of its properties that come from the various polymorphs, crystallographic defects and impurity defects has led to a lot of debate and many discoveries.

ZnO and it doped and alloyed phases have found application in oxide-based transistors, particularly thin film transistors (TFT).[127-129] These oxides provide a stable, high mobility, controllably thin, and (in the right energy range) transparent materials.[130,131] They also find use in flexible displays because of lower processing temperature compared to other amorphous TFTs, e.g. hydrogenated amorphous Si.

More recently, quasi-superlattices (QSL) of ZnO and related materials in with periodically heterojunctioned interfaces have shown spatially varying conduction band characteristics.[132] These QSL heterostructures also allow carrier mobility to be enhanced especially at the interface between the two oxides which also results in better Hall mobilities than either of the oxides alone. One pertinent example is the $In_2O_3$-$Ga_2O_3$-$ZnO$-$Ga_2O_3$-$In_2O_3$ QSL,[133] with showed an n-type carrier mobility >40 cm$^2$ V$^{-1}$ s$^{-1}$ ascribed to an interfacial 2-dimensional electron gas formed at the interface. This phenomenon was also reported for ZnO/ZnMgO interfaces[134] and bilayered $In_2O_3$-ZnO TFTs, all of which were processed using solution-based preparation and mobilities exceeding 45 cm$^2$ V$^{-1}$ s$^{-1}$ are possible.[135]

Making porous thin films of various oxides, and understanding the factors the lead to porosity in thin films processed from solution[136-138] is potentially useful for some applications. For applications in optoelectronics or related field, the porosity should ideally be controlled so that the film exhibits thin film properties while limiting as much as possible adverse effects on conductivity, carrier mobility or dispersion, while providing an opportunity to vary the effective refractive index. One approach we investigated several years ago was a porous analog of quasi-superlattice films, whereby dense and porous layers were iteratively grown in a periodic multilayer format. The objective was to form these porous layers while maintaining high crystal quality of a thin film structure.

Our solution-based approach to ZnO QSL involved iterative spin-coating with specific drying and annealing protocols to created periodic bilayered of dense and porous ZnO in thin film format with very high crystal quality for a porous layer. Near homoepitaxial growth was possible on $SiO_2$-coated silicon substrates at a maximum temperature of 300 °C that was c-plane oriented. The QSL morphology and some growth details are shown in Fig. 5. The QSL number can also be controlled, allowing thickness control with up to 20 layers investigated.[139,140] All of these structures exhibit classical ZnO photoluminescence with a pronounced defect sub-gap emission. The dense ZnO portion of each bilayer of the QSL is formed during the pre-anneal drying stage, and the cumulative annealing time overall crystallized the thick multi-layered SQL to strongly c-axis oriented overall QSL structure.

The types of films can also be doped from solution processing and we also demonstrated Al-doping to form QSLs of AZO. The process is near-identical, allowing control of the number of layers that form the QSL with a similar bilayered structure of dense and porous AZO with high c-axis crystallinity throughout.[141] ZnO QSL are solution processed using zinc acetate dihydrate [$Zn(CH_3COO)_2 \cdot 2H_2O$] dissolved in 2-methoxyethanol [$CH_3OCH_2CH_2OH$] and stabilized with the equimolar addition of monoethanolamine. To add an Al dopant, an Al-solution was prepared using aluminium nitrate-nonahydrate [$Al(NO_3)_3 \cdot 9H_2O$] in 2-methoxyethanol.[142] Figure 5 shows the resulting structures and their similarity to the ZnO layers. These QSL structures were investigated for their reflection properties and to assess the degree of dispersion from the porosity within the thin films that have a lower effective refractive index. Directional reflectance for single layers or homogeneous QSLs of ZnO and AZO channel materials exhibit a consistent degree of anti-reflection characteristics from 30-60° (~10-12% reflection) for thickness ranging from ~40 nm to 500 nm. The reflectance of AZO single layer thin films is <10% from 30 - 75° at 514.5 nm, and <6% at 632.8 nm from 30-60°. The data show that ZnO and AZO with granular or periodic substructure behave optically as dispersive, continuous thin films of similar thickness, and angle-resolved spectral mapping provides a design rule for transparency or refractive index determination as a function of film thickness, substructure (dispersion) and viewing angle.

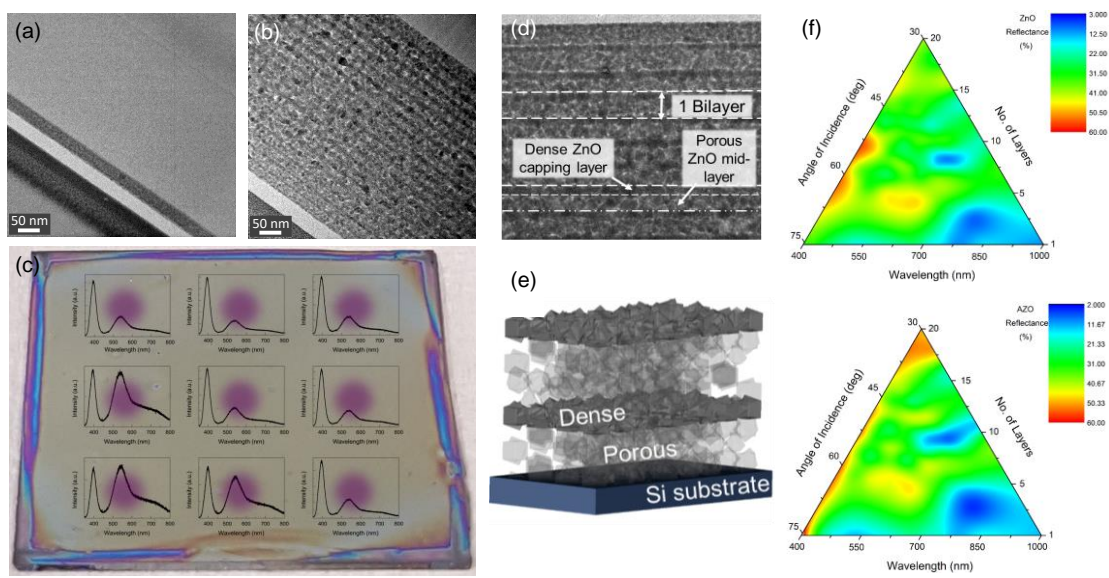

**Figure 5.** (a) TEM image of a cross-sectional FIB sectioned lamella of a single ZnO layer on oxidized silicon and (b) of a 20 QSL ZnO layer film. (c) Optical image of a QSL spin-coated on SiO$_2$/Si and corresponding PL spectra from 9 regions of the sample surface. (d) Representation of the porous and non-porous regions of a QSL layer also shown schematic in (e). (f) Reflectance intensity mapping as a function of angle of incidence and multilayer thickness (number of layers) in the UV–vis–NIR regions for full sample sets (1–20 layers) of ZnO and AZO QSL thin films.

*Porous Indium Tin oxide (ITO)*

Making porous transparent conducting oxides[143-145] for optoelectronic devices provides the option for graded index layers that reduce reflection losses at index mismatched interfaces. One way to form a porous layer bottom-up is to grow a nanowire layer with side branching. In this manner, the NW density is lower in the portion of the layer that is further from the substrate. We developed a controllable molecular beam epitaxial growth system for obtaining high quality, single phase, branched ITO NWs on silicon and oxidised silicon surfaces using In and Sn precursors in an oxygen atmosphere.[10,146-151] Briefly, The NW layer morphology is achieved by controlling the evaporation-condensation parameters during molecular beam epitaxy. It results in a bottom-up grown layer of self-catalysed and seeded NWs over large area (several cm$^2$) substrates. The ITO NW crystallographic quality is essentially defect-free, and its branched structure and high areal density of initial seeds resulted in a graded degree of porosity and thus refractive index. Figure 6 shows the growth summary, composition, crystallinity and electron images of an ITO NW layer.

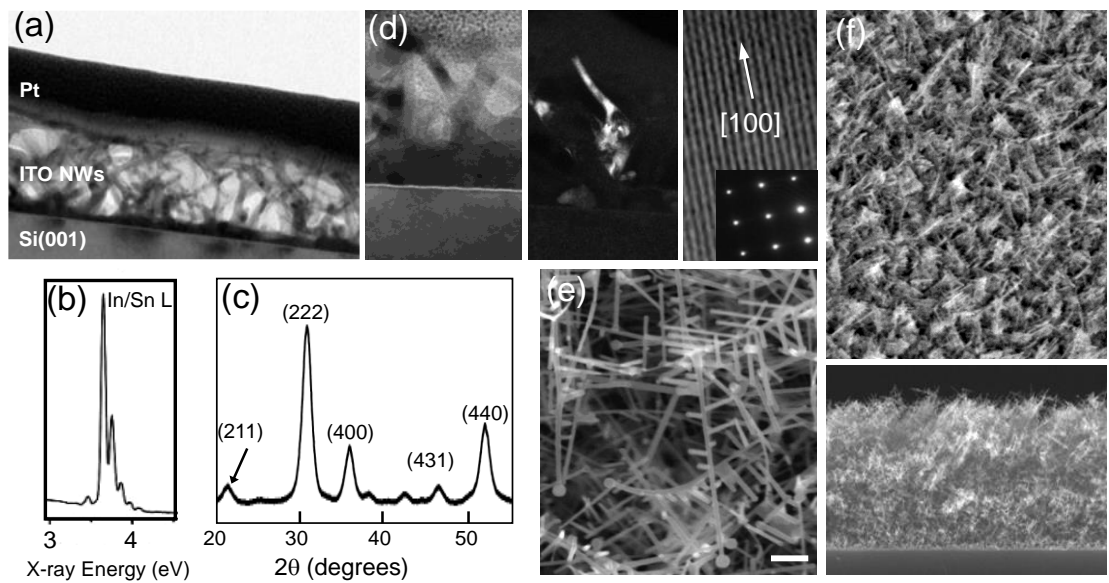

**Figure 6.** TEM image of a region of the ITO NW layer at early stages of NW growth. (b) EDX showing the In and Sn L-edge lines, O and Si substrate lines of $(In_{1.875}Sn_{0.125})O_3$. (c) XRD pattern for early stages of NW growth indexed to the $(In_{1.875}Sn_{0.125})O_3$ crystalline structure. (d) Bright and darkfield TEM images of NW growth from the In-Sn-O nucleation seeds. (e,f) SEM images of the ITO NW layer in side-view, plan view and at higher magnification inside the layer.

Porous ITO layers offer the ability to improve on the Fresnel reflection losses at the interface between the higher index ITO and (typically) air surrounding the layer. As the porosity and effective index are graded so that a lower refractive index is nominally at the air interface, we measured the reflectance of these ITO NW layers. For a given angle, the internal optical scattering and absorption were found to depend on the NW size, distribution, branching and overall morphology. The reflectivity at near-normal incidence (10°) shown in Fig. 7 was measured for the ITO NW layer, and a comparative porous ITO layer, and a dense thin ITO film. Finally, reflectivity of a standard antireflection layer of ITO on glass was measured to benchmark all ITO layers. ITO NW layers on silicon show significantly lower reflectivity than dense ITO on silicon or porous ITO layers. Only antireflection ITO coatings on glass showed similar transmission above 850 nm, and a usefully low sheet resistivity between ~15-65 Ω sq$^-$

[1], similar to values obtained from silver NW networks[152-156] and others based on carbon nanostructures optoelectronic films. Fresnel reflection associated with higher refractive index ITO thin films is virtually eliminated because of impedance matching at the ITO NW-air interface. We measured a drop from ~2.2 for the denser NW layer at the substrate interface to 1.04 at the air interface.

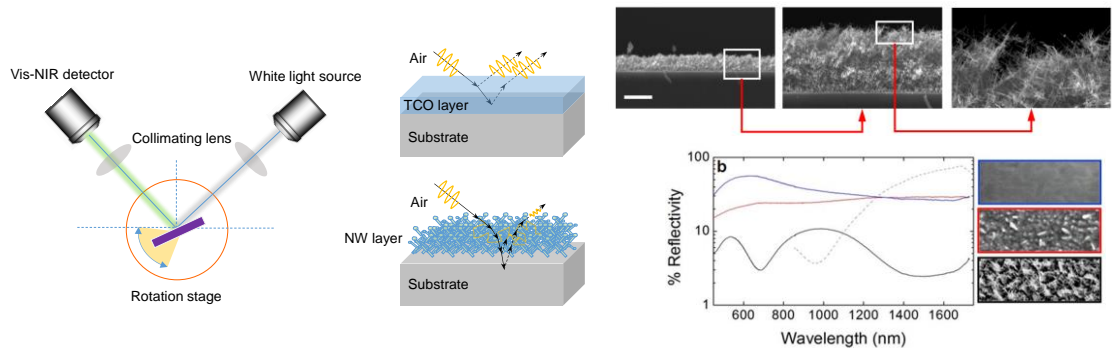

**Figure 7**. (Left) Schematic of the angle-resolved reflectance setup for measuring ITO NW layer properties. (Right) (a) A typical ITO NW layer grown on Si(100). The layer comprises a high density of branched NWs, which grow from a near compact contact layer to Si, followed by increased porosity from lower density growth away from the substrate. (b) Reflectance spectra for (——) a thin ITO antireflection coating, (——) a porous ITO layer, (——) an ITO NW layer (all on silicon) and (------) a thin ITO film on glass. Plan view SEM images of each layer are also shown.

ITO is more metal-like at lower photon energies in the NIR region. A Burnstein-Moss shift together with coupling to plasmons limits transmission in these regions. However, the graded porosity ITO NW layer (Fig. 8(a)) shows ~500 nm shift of the plasma frequency compared to a commercial ITO transparent oxide thin film, shown in Fig. 8(b). Angle resolved measurements of the reflectance in Fig. 8(c) further from the near-normal incidence of 10° to 75° at energies further into the infrared (2.5 µm) wavelengths. The graded ITO NW layer show maximum transmission at the 1.55 µm telecommunications energy window and does so at near normal incidence, a property that proved useful when maximizing light output from IR region LEDs and other technologies where light output light cone is most intense normal to the surface.

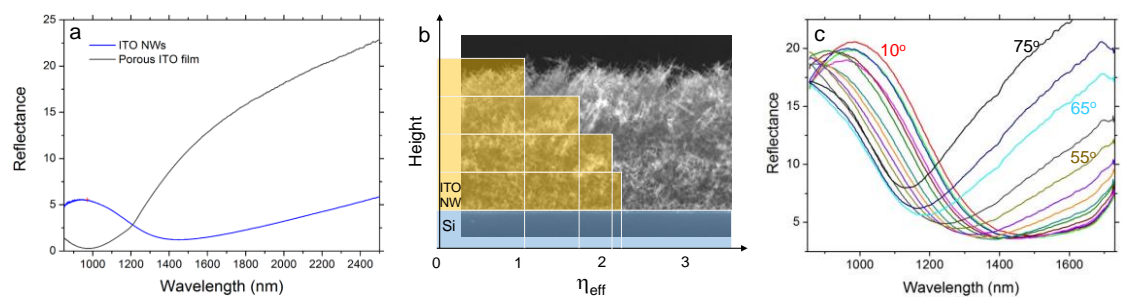

**Figure 8.** (a) Comparison of the transmission of an ITO NW layer and an MBE grown ITO thin film on glass at normal incidence. (b) Variation of effective refractive index determined from spectroscopic ellipsometry and the Bruggeman effective medium approximation at several stages of NW growth at 1.55 µm. (c) Angle resolved reflectance measurements of the ITO NW layer shown in (b).

Porosity is one of the most effective ways to release refractive indices less than ~1.2, a necessary objective for a wide variety of optical devices involving photonic crystals, broadband absorbers, omnidirectional reflectors, and better antireflection coatings with degrees of spectral transparency that can be controlled by material growth.

*Inverse opals*

High performance Li-ion batteries (LIBs) are one way to provide high capacity or energy density, with good, long life cycling stability at high current density.[157-159] To achieve this, new materials with properties that improve ion and electron transport kinetics with adverse side reactions are always important, particular those that are sustainable. Controlling material size, arrangement, surface chemistry and the nature of the reaction process with lithium are common strategies. Similar requirements are needed for other applications where optical properties and surface area play a significant roles, such as in photocatalysis and optical sensors, and for real-time diagnostics of various materials' behavior.

One approach is to fashion a three-dimensional (3D) interconnected network which is often called an inverse opal (IO)[160,161], and examples of which are shown in Figs 9-11. In response to this challenge, bicontinuous porous electrodes were proposed where the electrode active materials were put on pre-prepared conductive inverse opal structured nickel.[162,163] These types of electrodes were promising in the early days[164], showing higher initial capacities and quite remarkable rate response in battery cells.

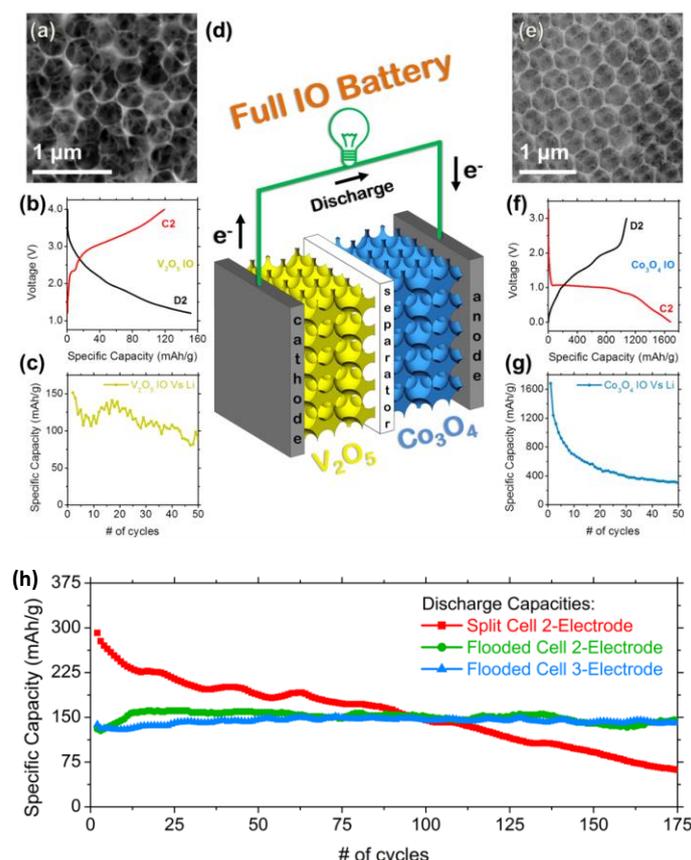

**Figure 9.** (a-g) The behaviour of $V_2O_5$ IO cathodes and $Co_3O_4$ IO anodes in lithium half cells. (h) A comparison of the discharge capacities for $V_2O_5$ IO/$Co_3O_4$ IO cells in a 2-electrode split cell and 2 and 3-electrode flooded cells.

We carried our investigations over several years with inverse opals of transition metal oxides for lithium reactions in batteries, photocatalysis and other applications. $TiO_2$ IOs have also been reported where capacity is retained a >98% efficiency for at least 5000 cycles.[165] While volumetric swelling occurs during lithiation, these IO retain their interconnectivity during long term cycling and provide surprising stability, capacity retention, and cycle life without the typical additives used in a Li-ion battery melectrode slurry, such as polymer binders or conductive carbons.

Battery researchers have for a several years considered using materials with so-called functional porosity that wets a greater volume of electrolyte to a lot of the active material with small dimensions, to remove solid state cation diffusion limitations especially for faster charging rates.[166-169] Random and organized pores in materials have been analysed and compared in great detail. The periodically arranged porous 3D material architecture[170] in IOs provide a defined material dimension and internal pore volume and overall gravimentric and volumetric mass loading per unit area compared to a compressed random material composite when investigating the effects of pores on electrochemical response.[171]

One example that demonstrates the half cell response of IO cathode and anodes, as well as an all-IO structures battery cell is that of $V_2O_5$ paired with $Co_3O_4$, using intercalation reaction at the positive electrode and conversion reactions at the negative electrode[172]. This investigation followed from separated tests of these materials individually,[173-176] the paired full-IO battery shown in Fig. 9 was the first demonstration of an all ordered porous material anode/cathode system that balanced the response of each electrode.

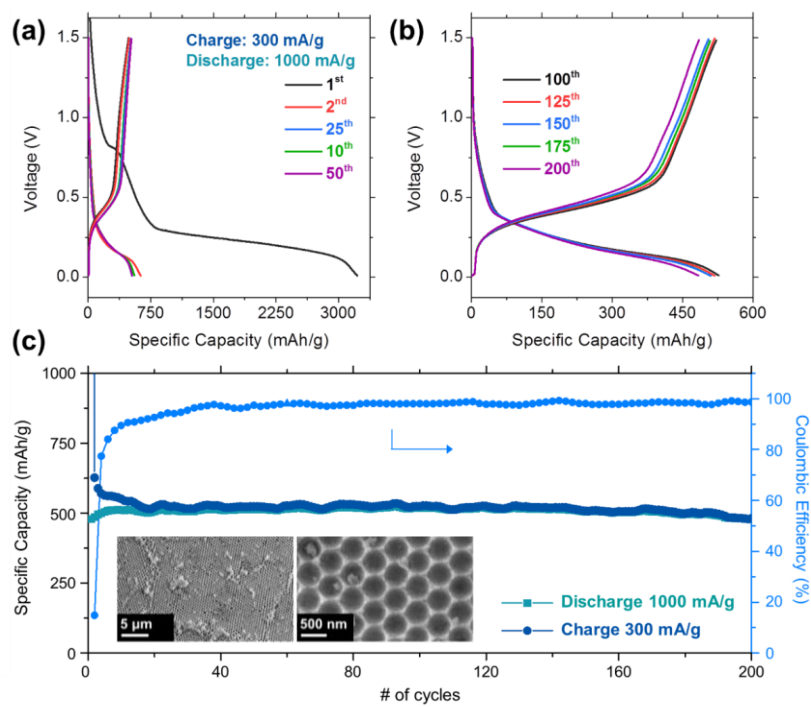

**Figure 10.** (a) Rate response of a $GeO_2$ IO over 100 cycles from 250 to 1000 mA/g. (b,c) Charge and discharge voltage profiles for charging at 300 mA g$^{-1}$ and discharging at 1000 mA g$^{-1}$ between 1.50 – 0.01 V. (d) Specific capacity and coulombic efficiency from asymmetric cycling. Inset: SEM images of the $GeO_2$ IO.

On the anode side, Germanium based materials are also interesting since, like silicon, it is a metalloid with a high theoretical capacity (1384 mAh/g) and it can be readily processed in nanoscale form much like silicon.[177-180] Using $GeO_2$ as the initial electrode formulation may provide opportunities for Ge-based anodes. In IO structure, the material has the inbuilt porosity that is often shown to be beneficial for high capacity anode materials. A Ge nanocrystal-containing $GeO_2$ inverse opal Li-ion battery anode was created from a germanium (IV) ethoxide ($Ge(OC_2H_5)_4$) precursor that was able to provide high capacity at good rates over 200 cycles. Some details are shown in Figure 10. Even under asymmetric galvanostatic discharging at 1000 mA g$^{-1}$, which was 3× faster than the charging rate, the GeO$_2$ IO maintained charge capacities of 524 and 508 mAh/g after the 50$^{th}$ and 150$^{th}$ cycles.

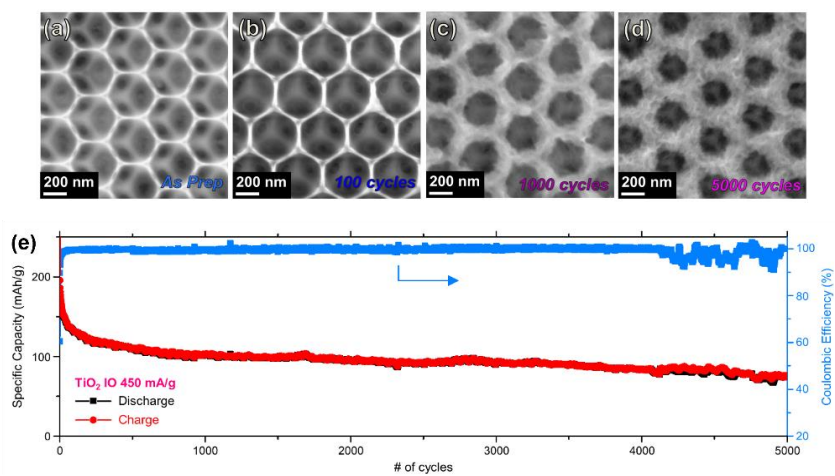

**Figure 11.** (a-d), SEM images of the as-prepared TiO$_2$ electrode, and after 100, 1000 and 5000 cycles, showing lithiation-induced swelling. (e) cycling behavior of a rutile TiO$_2$ Li-ion battery anode over 5000 cycles at a rate of 450 mA g$^{-1}$.

Until recently, the long term cycling behaviour of IO structured TiO$_2$ was not known. McNulty et al. tested these forms of anodes in Li-ion half cells and reported long-term cycle life behaviour over a 5000 cycle test at 450 mA g$^{-1}$, as shown in Fig. 11. The stability is capacity retention of long durations (5000 cycles) was due to the interconnected structure in spite of the change in material composition and conductivity during reversible lithiation.

We demonstrated this capability and the effect of band alignment with Au NP-immobilised TiO$_2$ and V$_2$O$_5$ IO materials for the photocatalytic hydrogenation of 4-nitrophenol under broadband and monochromatic illumination. Figure 12 shows the IO materials system.

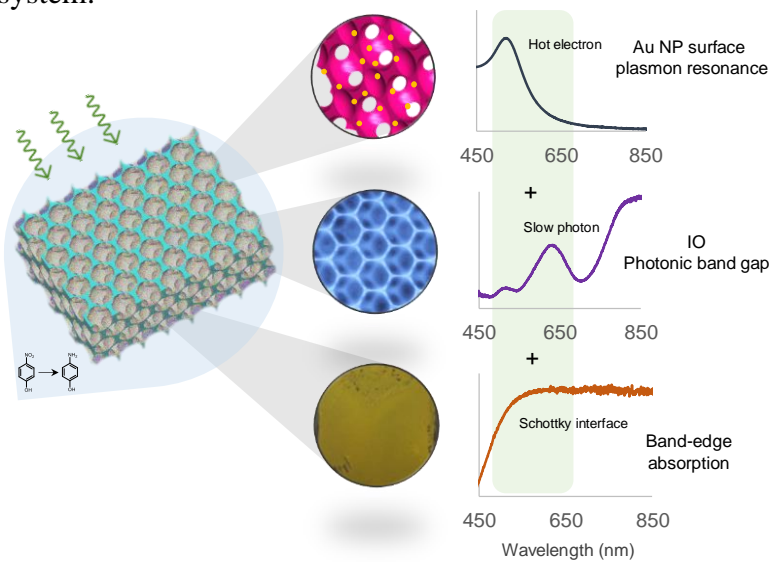

**Figure 12.** Representation of semiconductor photonic crystal plasmonic photocatalyst design and synergy. Inverse opal photonic crystals functionalised throughout their surface with Au NPs photocatalytically reduce nitrophenol under green (532 nm) laser light or broadband white light.

The highest catalytic enhancement was achieved when Au NPs were deposited on a visible light responsive semiconductor, $V_2O_5$ IO catalyst under green light excitation due to spectral overlap of the electronic bandgap, LSPR and excitation source. This was compared to IO $TiO_2$ and to both materials deposited as a thin film with immobilised Au NPs. By band alignment to enable charge transfer to the Au NP under optical pumping from the conduction band of the semiconductor, at a similar energy to the surface plasmon resonance and the maximum absorption of the photonic crystal of the material, catalysis of the hydrogenation reaction was significantly improved even under green light illumination as summarised in Fig. 13.

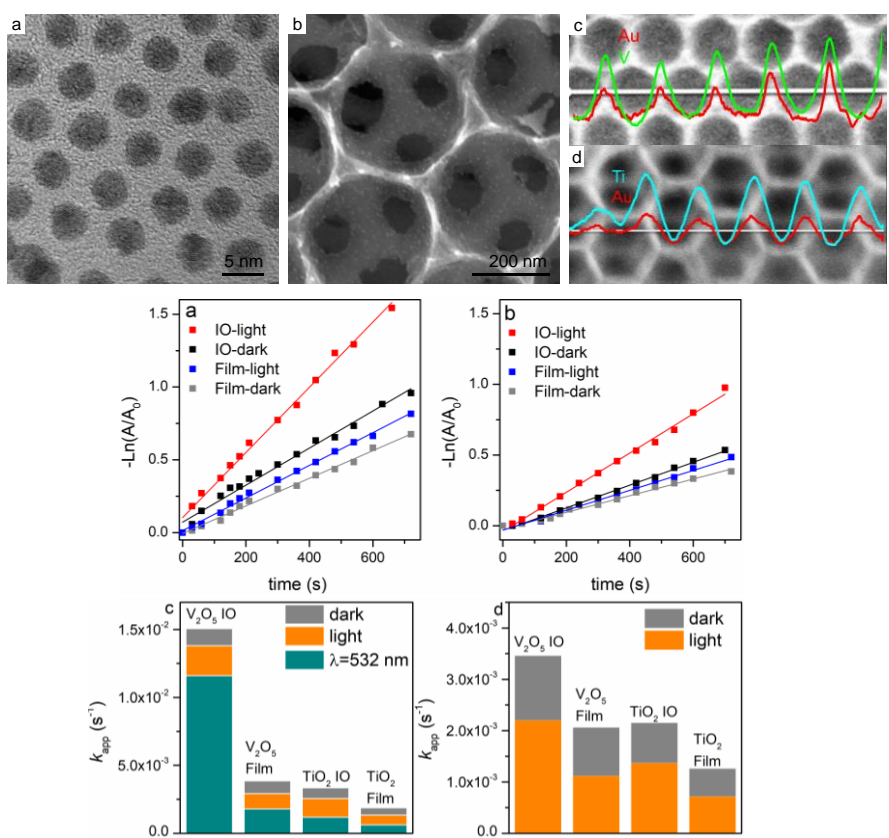

**Figure 13.** (Top) (a) TEM image of synthesized crystalline Au NPs. (b) $V_2O_5$ IO after Au NP immobilization on all surfaces throughout the 3D periodically porous structure. (c) EDX line scan of $V_2O_5$ and (d) $TiO_2$ IO photocatalysts decorated with immobilized Au NPs. (Bottom) Reaction profiles for plasmonic photocatalytic 4-nitrophenol reduction using IO and non-porous (non-p) thin film (a) $V_2O_5$ and (b) $TiO_2$ catalysts in the dark and under visible light and monochromatic (green) irradiation.

The modularity of the ordered porous semiconductor approach facilitates rational design of efficient plasmonic photocatalysts as the nanoparticle and semiconductor components can be readily altered, enabling it to be extended to other metal–semiconductor composites for a variety of catalytic applications.

*Porous 3D printed materials*

Metals foams[184] are randomly porous interconnected structures that can be made from stainless steel, nickel, gold and other alloys and compounds by various methods including templating, dealloying, electrodeposition, investment casting or deformation forming etc. They are often used in filtration and as support materials for other coatings or fillers. In electrochemical technologies, dense porous foams are very important in fuel cells, water splitting for hydrogen production, energy storage devices such as supercapacitors and in some cases batteries, and in other application in mechanical engineering. When the porous metal is structures with periodic ordered, they are often referred to as microlattices,[185-187] and these can typical exhibit some unique and impressive mechanical properties.

Porous metallic materials can be useful in electrochemical technologies as supports, porous conductors with overlayers of active materials or as active electrodes in their on right either that is intrinsic to the high surface area of the porous metal.[188] This intrinsic activity is sometimes beneficial, but for applications here the porous metal acts as a benign support, protective coatings or other ways of ensuring passive behavior are often required.

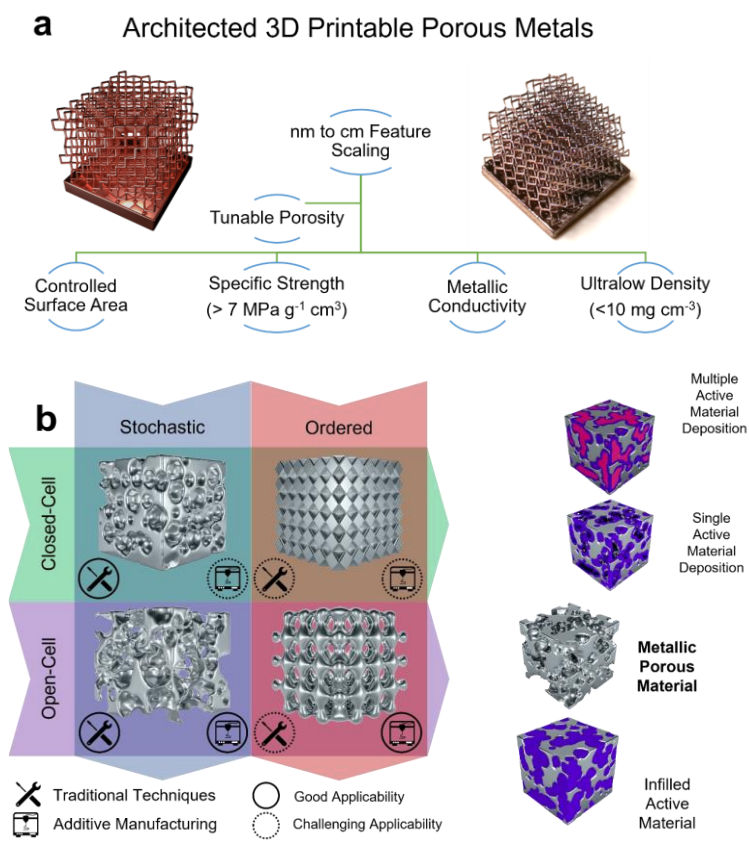

**Figure 14.** (a) 3D printed porous metals and associated characteristics by metallization of photopolymerized resins using stereolithographic printing. (b) Various porous metal geometries that can be produced by 3D printing. Filling or coating these types of printed porous metals with active materials can provide a range of unique active material formulations in almost any structure.

3D printing of porous metals opens the possibility for complete control over electrode design prior to manufacture and use, and made into exactly the complex structure slated for the particular application. We have been developing 3D printing method using stereolithographic photopolymerization that allows complex microlattice current collector electrodes to be formed with porosity designed for the application requirements.[189,190] Figure 14 shows some typical example or metal-coated 3D printed lattices and an overall schematic of the types of porous metals and geometries that can be formed. By coding and printing porosity with different degrees of order, pores sizes, geometries etc., these effects can be tested and optimized and applied to many technologies where the porous structure needs to be ceramic, metallic, carbonaceous, polymeric or an alloy or more complex inorganic compound. 3D printing offers reproducibility to make very complex pore systems, and give control of scaling of pore periodicity, shape, dimension, fill factor, tortuosity etc. that can be inherently challenging through coupled self-assembly and infilling, dealloying or etching, or electro(less)deposition and related methods.

## Conclusions

Porous materials, especially porous semiconductors, offer a very wide range of interesting properties compared to the bulk counterparts, that can allow different ways of exploiting their semiconducting nature. Many electronic material semiconductors can find application in optics and sensing based on exploitation of the electronic characteristics but also the properties made possible by lower refractive index, roughened or periodic porosity. Periodic porosity provide a photonic band gap and thus a method to selectively tune the energy range for transmission or reflection, random porosity and roughness can markedly influence heat conduction without adversely degrading electronic conductivity, useful for thermoelectric applications and may others. And many of the oxides of transition metal used so often in energy storage can be leveraged for other electrochemical technologies such as photocatalysis that would benefit from enhanced or tunable light absorption in several properties of the materials (photonic bandgap, electronic bandgap, surface plasmon resonance etc.). Porous version of semiconducting materials can provide opportunities for those materials use in application that otherwise may not have been considered. And or course, when you venture past semiconductors, there is a universe of possibilities and exotic properties and applications that await when designing and engineering pore structures for mechanical, biomedical, structural and other reasons.

## Acknowledgements

Support from several agencies is acknowledged including Enterprise Ireland, Irish Research Council, Science Foundation Ireland, EU FP6, FP7, and Horizon 2020 programmes.